\documentclass[debug]{rmaa}  
\usepackage{graphicx,subfigure,times,color}
\usepackage{amsmath}
\usepackage[colorlinks]{hyperref}
\hypersetup{colorlinks,citecolor=blue,linkcolor=blue,urlcolor=blue}  

\title{The influence of stellar objects mass distribution on their gravitational fields}

\author{
Vladimir Stephanovich,\altaffilmark{1} 
W{\l}odzimierz God{\l}owski,\altaffilmark{1}
Monika Biernacka, \altaffilmark{2}
and B{\l}a{\.z}ej Mrzyg{\l}{\'o}d, \altaffilmark{1}}

\altaffiltext{1}{Uniwersytet Opolski, Institute of Physics, ul.  Oleska  48,
	45-052 Opole, Poland.}
\altaffiltext{2}{Institute of Physics,  Jan Kochanowski University, ul. Swietokrzyska 15, 25-406 Kielce, Poland.}

\shortauthor{Stephanovich, God{\l}owski, Biernacka, Mrzyg{\l}{\'o}d}
\shorttitle{Mass distribution}

\fulladdresses{
	\item Uniwersytet Opolski, Institute of Physics, ul.  Oleska  48,
	45-052 Opole, Poland (stef@uni.opole.pl).
	\item Uniwersytet Opolski, Institute of Physics, ul.  Oleska  48,
	45-052 Opole, Poland (godlowski@uni.opole.pl).
	\item Institute of Physics,  Jan Kochanowski University, ul. Swietokrzyska 15, 25-406 Kielce, Poland.
	\item Uniwersytet Opolski, Institute of Physics, ul.  Oleska  48,
	45-052 Opole, Poland.} 

\listofauthors{V.A. Stephanovich, W. God{\l}owski, M. Biernacka, B. Mrzyg{\l}{\'o}d}
\indexauthor{Stephanovich, V. A.}
\indexauthor{God{\l}owski, W.}
\indexauthor{Biernacka, M.}
\indexauthor{Mrzyg{\l}{\'o}d, B.}
\abstract{
We study the influence of the astronomical objects masses randomness on the 
distribution function of their gravitational fields.  Based on purely theoretical arguments and comparison with
extensive data, collected from observations and numerical simulations, we have shown that while mass 
randomness does not alter the non-Gaussian character of the gravitational fields distribution, it changes 
the dependencies of mean angular momenta of galaxies and clusters on their richness. The specific form of 
above dependence is determined by the interplay of mass distribution character and different assumptions 
about cluster morphology. We trace the influence of masses distribution on the time evolution of stellar 
objects angular momenta in CDM and $\Lambda$CDM models. Our theoretical predictions are in very good coincidence 
with the statistical results derived both from observational data and numerical simulations.}
\resumen{}

\addkeyword{Galaxy:General}
\addkeyword{Galaxy:Formation}

\begin{document}
\maketitle

\section{Introduction}

As the gravitational fields are highly nonuniform during galaxies and their clusters formation, the distribution
of former plays an important role. Moreover, the character of gravitational field distribution permits to discern 
the specific scenario of galaxies formation. The classical scenarios of such formation had 
been proposed quite long time ago \citep{peb69,Zeldovich70,Sunyaew72,Doroshkevich73,Shandarin74,Silk83,Dekel85}
and dealt primarily with so-called Zeldovich pancake model \citep{Zeldovich70}, based on large gravitating 
body (of the pancake shape, hence the name) mechanical instability without any randomness in gravitational 
fields of the constituents \citep{Zeldovich70, shanzel89, long08}. These classical scenarios have not lost 
significance till now as new scenarios are essentially modifications of the classical ones and can be classified 
according to them. The improved form of the above classical scenarios has been put forward more recently, 
see \citep{Shandarin12,Giah14} for relevant references. The presence or absence of the random gravitational 
field fluctuations contribute to the studies of galaxies angular momentum acquisition during their formation stages. 
This may in principle permit to specify the most probable (among many other) scenario of the large stellar 
objects emergence. This is because the final test of a given scenario correctness is the comparison of its predictions 
with observations. The investigations of the variations of structures angular momenta gives us the 
opportunity to do so. Note that different scenarios make different predictions 
about galaxies orientations, i.e. their angular momenta alignments in structures 
\citep{peb69,Doroshkevich73,Shandarin74,Silk83,cat96,Li98,Lee00,Lee01,Lee02,Navarro04,Trujillio06,Zhang13}.
For this reason the analysis of galaxies’ planes orientation is regarded as a standard test of different
scenarios of the cosmic structures formation \citep{Rom12,Joachimi15,kiess15}.

One of the natural sources of gravitational fields randomness is the masses distribution 
of stellar objects. The simplest possible model of the masses distribution dates back 
to \citep{chan}, where the {\em{a priori}} given distribution function of masses 
$\tau(M)$ has been considered. In this paper, all characteristics of the stellar 
ensemble has been expressed through the different average powers of mass. 
The mass averaging had there been performed implicitly with the above function  $\tau(M)$. 
In principle, such approach can be generalized to the averaging over quadrupole and higher 
multipole moments of galaxies in the spirit of the \citep{apj15}. Although this effect may 
change some of the results quantitatively, we speculate that its overall influence will be rather faint. 
Next step has been done by \citep{shex74}, who considered the distribution function of stellar 
objects masses within the model of self-similar gravitational condensation. In this work,
 within the model of the expanding Universe in Friedmann cosmology, the stellar ensembles
 had been represented as a "gas" of self-gravitating masses, which can condense into aggregates 
with larger mass, forming finally very large clumpy objects. This model permits to derive the 
distribution of masses in the form  
\begin{equation} \label{yt1}
f(m)=A\left(\frac{m}{m_*}\right)^\alpha e^{-m/m_*},
\end{equation}
where $A$ is a normalization constant, see below. 
Note that the explicit expression of Shechter function \eqref{yt1} has been listed in the 
work \citep{shex76} devoted to the luminosity distribution of galaxies. Below we will use 
function \eqref{yt1} for the calculation of the amended (on the mass distribution) 
distribution function of the gravitational fields and angular momenta. Using this function, 
we assume that the mass is proportional to the first degree of a luminosity: $m \sim L$. 
Below we give the arguments why possible nonlinearity $m \sim L^\gamma$ ($\gamma \neq 1$) 
will not change our results qualitatively.

In the present paper, we consider tidal interaction in the ensemble of galaxies and their clusters in a 
Friedmann-Lema{\^i}tre-Robertson-Walker Universe with Newtonian self-gravitating dust fluid ($p = 0$)
 containing both luminous and dark matter. The commonly accepted model of such Universe is spatially
 flat homogeneous and isotropic $\Lambda$CDM model. The clumpy objects like galaxies and their 
clusters are formed as a result of almost scale invariant Gaussian fluctuations 
\citep{Silk68,Peebles70,Sunyaew70}. This assumption is the base of the so-called hierarchical clustering model \citep{dor70,Dekel85,peb69}.  The models with non-Gaussian initial fluctuations have also been considered in \citep{bart04}. The non-Gaussian character of distribution function has been postulated there, rather than calculated. 
Such calculation has been presented in \citep{apj15,raa17}, where the non-Gaussian distribution of 
gravitational fields and momenta were calculated using method of \citep{chan}. Here we generalize 
this calculation considering the masses distribution \eqref{yt1}. Note that the calculations made 
in \citep{apj15,raa17} dealt with equilibrium situation only. To consider non-equilibrium situation, 
it is necessary to use the differential equations of Fokker-Planck type with so-called fractional 
derivatives \citep{gs1,gs2}. In this case we can begin with ubiquitous Gaussian distribution and 
arrive at a non-Gaussian one as a result of primordial, fast time evolution. After it, the slower 
evolution, dictated by the $\Lambda$CDM scenario, takes place. Note that recently time evolution 
of intrinsic galaxies alignments has been found by \citep{Schmitz18}.

In hierarchical clustering approach, the large clumpy structures form as a result of gravitational interactions 
between smaller objects. In other words, the galaxies spin angular momenta arise as a result of tidal interaction 
with their neighbors \citep{Sch09}.  Note, that in the present paper the angular momentum is the result of tidal 
interaction with the entire environment, which occurs via interaction transfer from close to distant galaxies, see below. 
That is to say that our approach is the generalization of \citep{sch, cat96,ca96,Lee02}, where the average tidal 
interaction with the entire environment has been considered. In the present work we perform the theoretical and 
statistical analysis of the influence of tidal interaction between astronomical objects on the larger 
(then initial constituents) structures formation. We have also performed the comparison of our model predictions 
with vast data arrays, derived from observational and numerically simulated data. It turns out that our theoretical 
results are in pretty good coincidence between 
the above observational and numerical data. Our theoretical model includes additional distribution of masses, 
obeying Shechter function \eqref{yt1}. It turns out, that the mass distribution \eqref{yt1} does not change 
our main result \citep{apj15,raa17} that in the stellar systems with multipole (tidal) gravitational interaction, 
the distribution function of gravitational fields cannot be Gaussian. The crux of the matter here is a long-range character 
of Newtonian (and derived multipole) interaction between stellar objects. Such character implies 
that distant objects (like galaxies, their clusters and even dark matter haloes) still "feel each other", 
which is not the case for Gaussian distribution. The derived non-Gaussian distribution function allows 
to calculate the distribution of virtually any observables (like angular momentum) 
of the astronomical structures (not only galaxy clusters but smooth component like halos, which mass 
dominate the total mass of a cluster, see \citep{kb12}) in any (linear or nonlinear) Eulerian approach.

The relation between angular momentum of the galaxy clusters and their masses has also been 
investigated observationally. It is not difficult to analyze the distribution of the angular momenta for the luminous matter. In real Universe, the luminous galaxies and their structures are surrounded by dark matter halos.
These halos are often much more extended and massive than luminous component of the structures.
Unfortunately, direct observation of dark matter halos and their angular momenta is much more complicated.
One should not forget, however, that there are observed correlations  between luminous and dark matter 
(sub)structures. It implies the certain dependence between dark matter halos and luminous 
matter (real galaxies) orientations \citep{Trujillio06,Paz08,Pe08,Bett10,Paz11,Kim11,Varela12}.
Recently, the results of \citet{Okabe18,Codis18} based on Horizon-AGN simulation shows the similar dependence.
This allows to conclude, that the analysis of angular momentum of luminous matter gives us also the
information about angular momentum of the total structure (i.e. that with dark matter halos).  As a result,  the analysis of 
the angular momentum of "real" (luminous) galaxies and their structures, is still important as a test for possible
structure formation scenario. Note, that investigations  of galaxies orientation in clusters are 
also very important for the  analysis of weak gravitational lensing, see \citet{heav00,Hey04,kiess15,apj15,Codis16} for more details.

As generally galaxy clusters do not rotate \citep{Hwang07,Tovm15}, the angular momentum of a cluster is primarily due to spins of member galaxies. Unfortunately, usually we do not know angular momenta of galaxies. So the orientations 
of galaxies are investigated instead (\citet{Opik70,h4}, see \citet{Rom12,g19} for present review),
as it is assumed that the rotational axes of galaxies are normal to their planes. 
Such assumption seems to be quite reasonable at least for the spiral galaxies. As a result, 
stronger alignment of galaxies in a structure means larger angular momentum of the latter.

The question is, if there are any relation between the alignment and mass of the structures.
General result of the previous papers is that there is no sufficient evidence for galaxies 
alignment in less massive structures like groups and poor clusters. However, we observe
the alignment of galaxies in rich clusters, see \citet{g2011a,g19} for review.
First results (\citet{Godlowski05,Aryal07}) were qualitative only.
Because of that, \citet{g10a,g2011b} investigated quantitatively the orientation of galaxies in the 
sample of 247 rich Abell cluster, using improved \citet{h4} method (see \citet{g19} for latest review).
In these papers, it was found that the alignment is present in the above sample. Moreover, galaxy 
orientation increased with numerousness of the cluster. However, the data was not sufficient 
both to resolve the question about the exact form of this relationship and for confirmation of
the hypothesis that the angular momentum of the structure increases with time. This is the reason that
we decide to extend our sample and compare observational results with those obtained from simulations.
 
\section{The formalism} 

Similar to the papers \citet{apj15,raa17} we consider here the quadrupolar (tidal) interaction of the stellar objects
\begin{equation}
{\cal H}=- G\sum_{ij}Q_im_jV({\bf r}_{ij}),\ 
V({\bf r})=\frac 12 \frac{3\cos^2\theta - 1}{r^3}, \label{star1}
\end{equation}
where $G$ is the gravitational constant, $Q_i$ and $m_i$ are, respectively, the quadrupole 
moment and mass of $i$-th object, $r_{ij}\equiv $ $|{\bf r}_{ij}|$, ${\bf r}_{ij}=$  ${\bf r}_{j}-$ ${\bf r}_{i}$ 
is a relative distance between objects while $\theta$ is the apex angle. The Hamiltonian function \eqref{star1} 
describes the interaction of quadrupoles, formed both from luminous and dark matter, see \citet{raa17} for details. 

To account for the mass distribution \eqref{yt1}, we begin with the expression for characteristic 
function $F(\rho)$ of the random gravitational fields' distribution \citep{apj15,raa17}. 
\begin{equation}\label{sa1}
F(\rho)=\int_Vn({\bf r})\left[1-\frac{\sin \rho E({\bf r})}{\rho E({\bf r})}\right]d^3r.
\end{equation}
In the spirit of the article of \citet{chan}, we rewrite the expression \eqref{sa1} in the form
\begin{equation}\label{sa2}
F(\rho)=\int_{V,m}n({\bf r},m)\left[1-\frac{\sin \rho E({\bf r},m)}{\rho E({\bf r},m)}\right]d^3rdm,
\end{equation}
where $n({\bf r},m)$ is the number density (concentration, proportional to probability, see below) 
of stellar objects (galaxies, their clusters and also dark matter haloes) at the position ${\bf r}$ 
with a mass $m$. As the average density at large scales can be well regarded as constant 
(slowly spatially fluctuating to be specific), see \citet{chan, shex74}, the number density
 $n$ in Eq. \eqref{sa2} can be safely considered to be spatially uniform, i.e. $n=n(m)$. 
In this case the expression \eqref{sa2} reads
\begin{equation}\label{sa2a}
F(\rho)=\int_{V,m}n(m)\left[1-\frac{\sin \rho E({\bf r},m)}{\rho E({\bf r},m)}\right]d^3rdm,
\end{equation}
where 
\begin{equation}\label{sa3}
E({\bf r},m)=E_0\frac{3\cos^2\theta -1}{r^4},\ E_0=\frac{1}{2}GQ, \ Q\approx mR^2
\end{equation}
is the quadrupolar field \citet{apj15}, $m$ is the mass of a stellar object 
(like galaxy or cluster) and $R$ is its mean radius. We take the function $n(m)\equiv f(m)$ 
in the form of the Shechter function \eqref{yt1}, where $m_*$ and $\alpha$ are adjustable parameters. 
We obtain the normalization constant $A$ from the condition (see Eq. (4) of \citet{shex74}) 
\begin{equation}\label{sa4a}
n=\int_0^\infty n(m)dm,
\end{equation}
where $n$ is our previous constant concentration \citet{apj15,raa17}. 
Note, that there is no problem to take any other dependence $n(m)$ which will not 
complicate our consideration a lot. As we mentioned above, here we, following  \citet{shex76}, 
assume that the luminosity is directly proportional to the first degree of a mass ${\tilde L} \sim m$. 
But there is no problem to consider the higher degrees in this relation like  ${\tilde L} \sim m^k$, $k=4$. 
In this case, the argument of the function \eqref{yt1} will be $m^k$ instead of $m$.

The explicit calculation gives
\begin{eqnarray}
n=A\int_0^\infty \left(\frac{m}{m_*}\right)^\alpha e^{-m/m_*}dm=Am_*\Gamma(1+\alpha). \Rightarrow A=\frac{n}{m_*\Gamma(1+\alpha)}. \label{sa5}
\end{eqnarray}
Here $\Gamma(z)$ is Euler $\Gamma$ - function \citep{abr}. Finally we have from \eqref{sa2a}
\begin{equation} \label{sa6}
F(\rho)=\frac{n}{m_*\Gamma(1+\alpha)}\int_V\int_0^\infty dm \left(\frac{m}{m_*}\right)^\alpha e^{-m/m_*} \left[1-\frac{\sin \rho E({\bf r},m)}{\rho E({\bf r},m)}\right]d^3r,
\end{equation}
where $E({\bf r},m)$ is given by the equation \eqref{sa3}. It turns out that the equation \eqref{sa5} 
can be reduced to the Eqs (17) and (18) from \citet{apj15} but with slightly renormalized 
coefficient before $\rho^{3/4}$. This is because under the assumption that $n$ does not depend 
on coordinates (it depends only on mass, see Eq. \eqref{yt1}), the coordinates and mass turn out 
to be effectively decoupled. To do so, we perform first the integration over $d^3r$ in \eqref{sa5}. 
This integration is exactly the same as that in \citet{apj15} 
(since the mass enters Eq. \eqref{sa3} through parameter $E_0$ which is unimportant for coordinate integration) 
so that we have from \eqref{sa6}
\begin{equation} \label{sa7}
F(\rho)=2\pi\cdot 0.41807255\rho^{3/4}E_{10}^{3/4}\int_0^\infty m^{3/4}n(m)dm, \ E_{10}=\frac{1}{2}GR^2.
\end{equation}
The integral in \eqref{sa7} can be performed as follows
\begin{eqnarray}
I=\int_0^\infty m^{3/4}n(m)dm=\frac{n}{m_*\Gamma(1+\alpha)}\int_0^\infty m^{3/4} \left(\frac{m}{m_*}\right)^\alpha e^{-m/m_*}dm=\nonumber \\
=\frac{n\ m_*^{7/4}}{m_*\Gamma(1+\alpha)}\int_0^\infty x^{\alpha+3/4}e^{-x}dx=n\ m_*^{3/4}\frac{\Gamma\left(\alpha+\frac 74\right)}{\Gamma(\alpha+1)}.
\end{eqnarray}
Finally 
\begin{equation}\label{sa8}
F(\rho)=2\pi \ n m_*^{3/4}\frac{\Gamma\left(\alpha+\frac 74\right)}{\Gamma(\alpha+1)}E_{10}^{3/4}\cdot0.41807255\cdot \rho^{3/4}\equiv \kappa \rho^{3/4},
\end{equation}
where 
\begin{equation}\label{sa9}
\kappa=2\pi \ n \cdot0.41807255\cdot E_0^{*3/4}\frac{\Gamma\left(\alpha+\frac 74\right)}{\Gamma(\alpha+1)},\ E_0^*\equiv m_*E_{10}=\frac 12 Gm_*R^2\equiv \frac 12 G Q^*.
\end{equation}
The expressions \eqref{sa8}, \eqref{sa9} give the answer for the case when we have Shechter distribution for galaxies masses. 
The difference between previous case (\citet{apj15,raa17}) of a single mass is that now the width of distribution function 
of random gravitational fields depends on the fitting parameters $m_*$ and $\alpha$.
Note that for {\em{any}} function $n(m)$ the result for characteristic function $F(\rho)$ will be 
Ex. \eqref{sa8} but with different coefficient $\kappa$.

\section{Calculation of the mass dependence of mean angular momentum}

To derive the mass dependence of mean angular momentum, we should first calculate 
the distribution function of gravitational fields $f(E)$, then using linear relation 
between angular momentum $L$ and field $E$ (here, without loss of generality, 
we consider the moduli of corresponding vectors, see \citet{apj15,raa17} for details), 
derive the distribution function $f(L)$, from which we obtain the desired dependence. 

The expression for the field $E$ distribution reads \citep{apj15}
\begin{equation}
f(E)=\frac{1}{(2\pi)^3}\int e^{iE\rho-F(\rho)}d^3\rho\equiv \frac{1}{2\pi^2 E}\int_0^\infty \rho e^{-\kappa \rho^{3/4}}\sin \rho E d\rho,
\end{equation}
where $F(\rho)$ is the characteristic function \eqref{sa8}. Function $f(E)$ is normalized as follows
\begin{equation}\label{zk1}
4\pi \int_0^\infty E^2 f(E)dE=1
\end{equation}
and coefficient $\kappa$ is given by the expression \eqref{sa9}. 
The distribution function of angular momenta can be expressed by usual way from above $f(E)$
\begin{equation}\label{zk2}
f(L)=f[E(L)]\left|\frac{dE(L)}{dL}\right|.
\end{equation}
This gives explicitly (see \citet{apj15,raa17})
\begin{equation}\label{zk3}
f(\lambda)=\frac{I(\lambda)}{2\pi^2\lambda^3\kappa^4L_0(t)},
\end{equation} 
where $\lambda=L/(L_0(t)\kappa^{4/3})$,
\begin{equation}
I(\lambda)=\int_0^\infty x\sin x \exp\left[-\left(\frac{x}{\lambda}\right)^{3/4}\right]dx.
\end{equation}
Function $L_0(t)$ defines the model (CDM or $\Lambda$CDM) used by \citet{raa17}. 

\subsection{CDM model}

Equation \eqref{sa8} shows that the distribution function of angular momenta for the case 
of distributed masses is similar to that from \citet{apj15} with the only change $\alpha \to \kappa$. 
This means that the mean dimensionless angular momentum reads \citep{apj15,raa17}
\begin{equation}\label{iu1}
\lambda_{max}=0.602730263.
\end{equation}
Now we should express parameter $\kappa$ through galaxy (but not cluster, which is 
still equals to $M=mN$, where $m$ is galaxy mass, $M$ is cluster mass, $N$ is a number 
of galaxies in a cluster, see \citet{apj15,raa17}) mass $m$. This quantity is now 
defined as an average mass with distribution \eqref{yt1}
\begin{equation}
m=\int_0^\infty m_1\ n(m_1)dm_1 \equiv \frac{n \  m_*^2\Gamma(\alpha+2)}{m_*\Gamma(1+\alpha)}=(\alpha+1)m_*n.\label{iu2}
\end{equation}
The expression \eqref{iu2} implies that the galaxy mass $m$ is related to the mass distribution parameter $m_*$ as
\begin{equation}\label{iu3}
m_*=\frac{m}{n(\alpha+1)}.
\end{equation}
The next step is to substitute the expression \eqref{iu3} to the expression \eqref{sa9} for $\kappa$ 
and express it through galaxy mass $m$ instead of $m_*$. We have from Ex. \eqref{sa9}
\begin{equation}
\kappa^{4/3}=\frac{mE_{10}\ n^{1/3}}{\alpha +1}\left[2\pi \cdot 0.41807255\cdot \frac{\Gamma\left(\alpha+\frac 74\right)}{\Gamma\left(\alpha+1\right)}\right]^{4/3}. \label{iu4}
\end{equation}
In CDM model, the function $L_0(t)$ has the form \citep{raa17}
\begin{equation}\label{ozh1}
L_0(t)=\frac{2I}{3}  \frac{t}{t_0^2},
\end{equation}
where $I \approx mR^2$ is a galaxy moment of inertia and $t_0$ is a time scale.
 We have from the equation \eqref{iu1} in dimensional units
\begin{eqnarray}
L_{max}=\lambda_{max}L_0\kappa^{4/3}\equiv 0.6027 \frac{2I}{3}  \frac{t}{t_0^2}\frac{mE_{10} n^{1/3}}{1+\alpha} \left[2\pi\cdot 0.418\cdot  \frac{\Gamma\left(\alpha+\frac 74\right)}{\Gamma\left(\alpha+1\right)}\right]^{4/3}\nonumber \\
 =0.7281884\frac{t}{t_0^2}\frac{m^2 n^{1/3}}{1+\alpha}\zeta^{4/3}GR^4,\ 
 \zeta=\frac{\Gamma\left(\alpha+\frac 74\right)}{\Gamma\left(\alpha+1\right)}.  \label{iu5}
\end{eqnarray}
The comparison of the expression \eqref{iu5} with Ex. (12) from \citet{apj15} shows that their only 
difference is other power of $n$. Namely, while latter Ex. (12) involves $n^{4/3}$, 
our expression \eqref{iu5} contains $n^{1/3}$. This is the consequence of the star masses distribution 
according to the Shechter function. We note also that the above mass distribution leaves the power of 
galaxy mass $m$ intact, i.e. both expressions involve $m^2$. This gives that in the first scenario 
(see below and Eq. (13) of \citet{raa17}) the dependence of $L_{max}$ on galaxy cluster mass $M=mN$ will 
be the same $\sim M^{5/3}$. At the same time, in the second scenario (Eq. (15) of Ref. \citet{raa17}) the 
dependence on $M$ will be $M^{1/3}$ instead of $M^{4/3}$.
We now derive the dependencies on $M$ within both scenarios of Ref. \citet{raa17}.

\subsubsection{First scenario}
In this scenario we represent galaxy volume as $V=R^3$ \citep{raa17}, where $R$ is 
the mean radius of a galaxy. In this case we have from \eqref{iu5}
\begin{eqnarray}
&&L_{max}=\eta m^2n^{1/3}R^4=\eta m^2R^4\frac{N^{1/3}}{V^{1/3}}\equiv \eta m^2R^4\frac{N^{1/3}}{R}=\eta m^2R^3N^{1/3}=\nonumber \\
&&=\eta \frac{R^3}{N} M^{5/3}\frac{m^{1/3}}{N^{1/3}}=\eta \frac{1}{n} M^{5/3}\frac{\rho^{1/3}}{n^{1/3}}=\eta M^{5/3}\frac{\rho^{1/3}}{n^{4/3}},\ \eta= \frac{t}{t_0^2}\frac{0.728G}{1+\alpha}\zeta^{4/3},\ \nonumber \\ 
&&\rho=\frac{m}{V},\ n=\frac{N}{V}.\label{irs1}
\end{eqnarray}
The comparison of Eq. \eqref{irs1} and Eq. (13) from Ref. \citet{raa17} shows that the $M^{5/3}$ is the same,
 but the galaxies concentration $n$ now enters in the power 4/3 instead of 1/3. One more difference is that 
now Shechter exponent $\alpha$ (see Eq. \eqref{yt1}) enters the answer via parameters $\eta$ and $\zeta$. 
It should be extracted from the best fit between expression \eqref{irs1} and data, taken either from observations 
or numerical simulations.
\subsubsection{Second scenario}
In this scenario the galaxy volume is $V=R_A^3$, where $R_A$ is a mean galaxy cluster radius. We have from \eqref{iu5}
\begin{eqnarray}
L_{max}=\eta m^2n^{1/3}R^4=\eta m^2R^4\frac{N^{1/3}}{R_A}=\eta \frac{R}{R_A}R^3m^2N^{1/3}=\nonumber \\
=\eta \frac{R}{R_A}R^3 m^2\frac{M^{1/3}}{m^{1/3}}=\eta \frac{R}{R_A}R^3 m^{5/3}M^{1/3}. \label{irs2}
\end{eqnarray}
 It is seen, that contrary to Eq. (15) of Ref. \citet{raa17}, here we have $M^{1/3}$. 
Also, the Shechter parameter $\alpha$ enters the answer.

\subsection{$\Lambda$CDM model}

Here, similar to \citet{apj15}, we should isolate the contribution from time dependent functions 
$f_{1,2}(\tau)$  ($\tau=t/t_0$, $t_0=2/(3H_0\sqrt{\Omega_\Lambda}$), see Ex. (46) of \citet{apj15} 
and $i=1,2$ numbers the orders (first and second respectively) of perturbation theory. Following \citet{apj15}, 
we have for argument of the distribution function $H(\lambda,t)$ (Eq. (44) of \citet{apj15})
\begin{equation}\label{sw1}
\lambda(\tau)=\frac{L}{I\kappa^{4/3}}\frac{1}{f_i(\tau)}.
\end{equation}
Similar to above CDM model, the maximum of the distribution function $\lambda_{max}=0.602730263$ generates following relation
\begin{eqnarray}\label{sw2}
L_{max}=I\kappa^{4/3}\cdot 0.6027f_i(\tau)\equiv \eta_{\Lambda CDM i}(\tau)m^2R^4n^{1/3},\nonumber \\ \eta_{\Lambda CDM i}(\tau)=f_i(\tau)\frac{0.6027\Psi^{4/3}G}{2(\alpha+1)},\ \Psi=2\pi \cdot 0.418\cdot \frac{\Gamma\left(\alpha+\frac 74\right)}{\Gamma\left(\alpha+1\right)}.
\end{eqnarray}
The relation \eqref{sw2} is almost similar both in CDM and $\Lambda CDM$ models. The only difference is in the 
functions $f_i(\tau)$, where in CDM model  
\begin{equation}\label{sw3}
f_1(\tau)=(2/3)\tau,\ f_2(\tau)=(-4/3)\tau^{1/3}.
\end{equation}
It is seen that the substitution of functions $f_{1,2}$ \eqref{sw3} yields immediately the expressions \eqref{irs1} 
and \eqref{irs2} for CDM model. 
In $\Lambda CDM$ model the functions $f_{1,2}(\tau)$ should be taken from the solution of the differential equations (30) 
and (31) of \citet{apj15}. 

As the expression \eqref{sw2} for $L_{max}$ is formally equivalent to Eqs \eqref{irs1} and \eqref{irs2}, 
the dependences $L_{max}(M)$ are the same as those defined by Eqs \eqref{irs1} and \eqref{irs2} except that we should use 
now  $ \eta_{\Lambda CDM i}(\tau)$.
\begin{figure*}
\begin{center}
\includegraphics*[width=0.9\columnwidth]{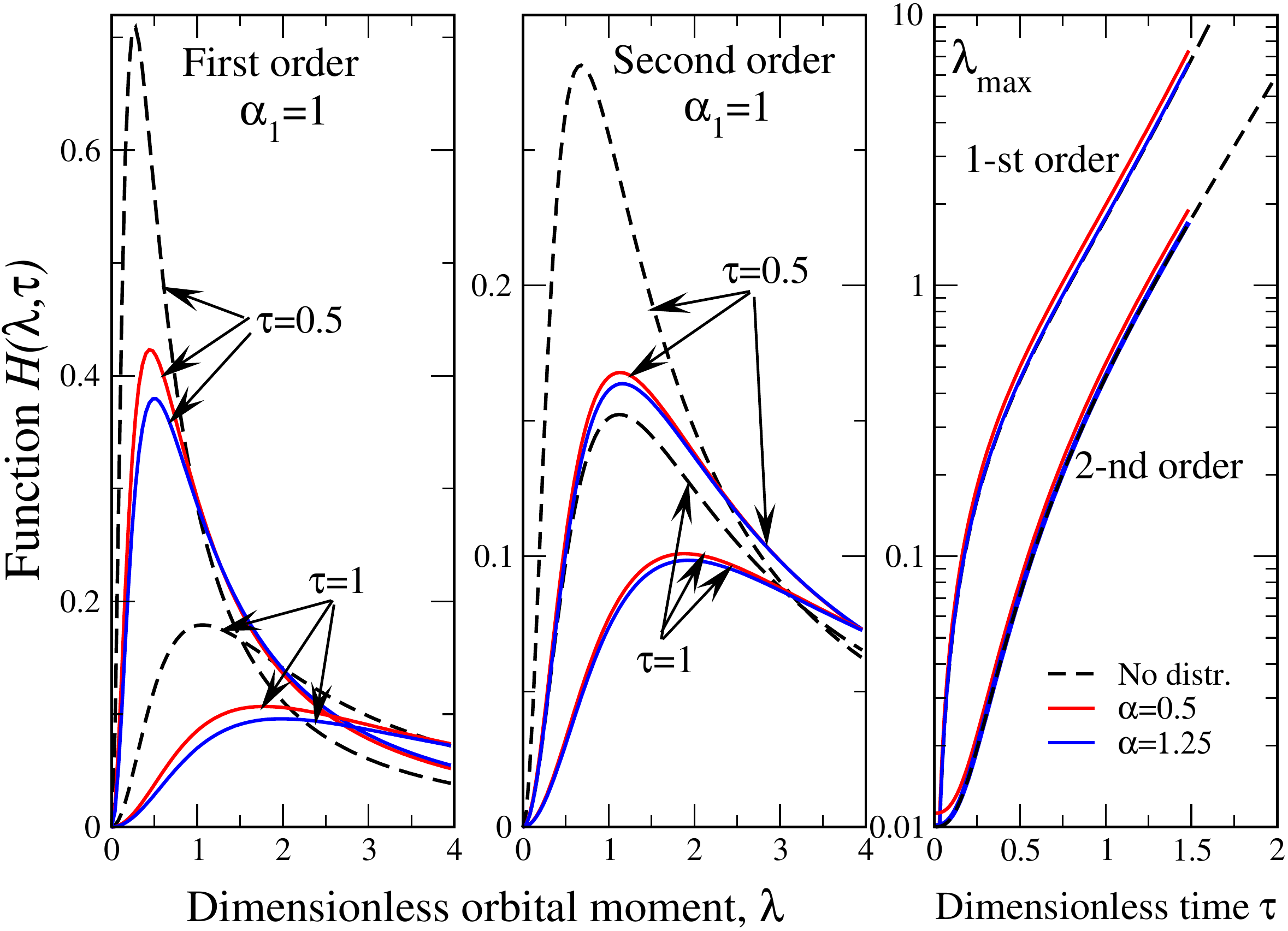}
\end{center}
\caption{Time evolution of the distribution function $H(\lambda,\tau)$ in $\Lambda$CDM model for the case of 
mass-dependent density \eqref{yt1}. Left panel: first order of perturbation theory. Middle panel: second order.
 Right panel shows the dependence $\lambda_{max}(\tau)$ in the first and second orders of perturbation theory. 
In all panels, dashed lines (marked as "No distr." in the right panel) correspond to the previous case of 
mass-independent density $n$. We consider two Shechter exponents $\alpha=0.5$ (red curves) and 1.25 (blue curves), 
coded by colors and explained in the legend in right panel. 
Parameter  of $\Lambda$CDM model $\alpha_1=\left(\frac{1-\Omega_\Lambda}{\Omega_\Lambda}\right)^{1/3}=1$.}
\label{fig:dis}
\end{figure*}

The representative plots of the function $H(\lambda,\tau)=2I(\lambda,\tau)/(\pi \lambda)$ \citep{apj15}
and the maximal value $\lambda_{\max}(\tau)$ for the mass-dependent density \eqref{yt1} are reported in 
Fig. \ref{fig:dis}. Left and middle panels show the results for the first (Eq. (30) of \citet{apj15}) and 
second orders (Eq. (31) of Ref. \citet{apj15}) of perturbation theory. First, it is seen, that the distribution 
of masses does not make qualitative difference in the shape of distribution function. Namely, the shape of 
the dashed black curves (those without masses distribution) and red and blue ones is the same. At the same time, 
the distribution functions $H(\lambda,\tau)$ has substantially smaller amplitudes in the case of mass distribution. 
This means that the presence of mass distribution changes the functions  $H(\lambda,\tau)$ quantitatively. 
The influence of Shechter exponent $\alpha$ \citep{shex74,shex76} is minimal - the curves $H(\lambda,\tau)$ 
as well as $\lambda_{max}(\tau)$ (right panel) are almost the same for $\alpha=1.25$ and $0.5$, which is a big difference. 
This means that while the distribution of masses by itself changes the distribution function quantitatively, 
the value of constant $\alpha$ in that distribution is of minute influence. Our analysis shows that the 
above tendency persists for any time instant (in Fig. \ref{fig:dis} we have only two time instants $\tau=0.5$ and 1) 
and any reasonable $\alpha>0$.
The maximal values $\lambda_{max}$ are almost independent of the presence of mass distribution. Really, it is seen 
from right panel of Fig. \ref{fig:dis}, that both curves for no mass distribution (black dashed lines) and those 
with it lie very close to each other. This means simply that the physics of the system under consideration is 
determined by the distribution of random gravitational fields,  than that of masses of stellar objects. 
Maybe the "complete" dependence $n({\bf r},m)$ \eqref{sa2} (rather than present simplified situation $n(m)$ 
in each spatial point ${\bf r}$ \eqref{sa2a}) will improve the situation. On the other hand, 
it is well acceptable that the robust distribution of gravitational fields is simply not susceptible to the 
small corrections like mass distribution. Latter, in turn, may mean, that next important step in the physics 
of galaxies formation is to consider the short-range interaction between galaxies (due to dark matter presence, 
for instance) so that the real average angular momentum (and not the distribution function maximum, considered so far) 
will appear, see Eq. (48) of Ref. \citet{apj15} and Eq. (28) of Ref. \citet{raa17}.

\begin{figure}
	\subfigure[Relation between $N$ and $\chi^2$]{\includegraphics*
		[width=0.45 \textwidth]{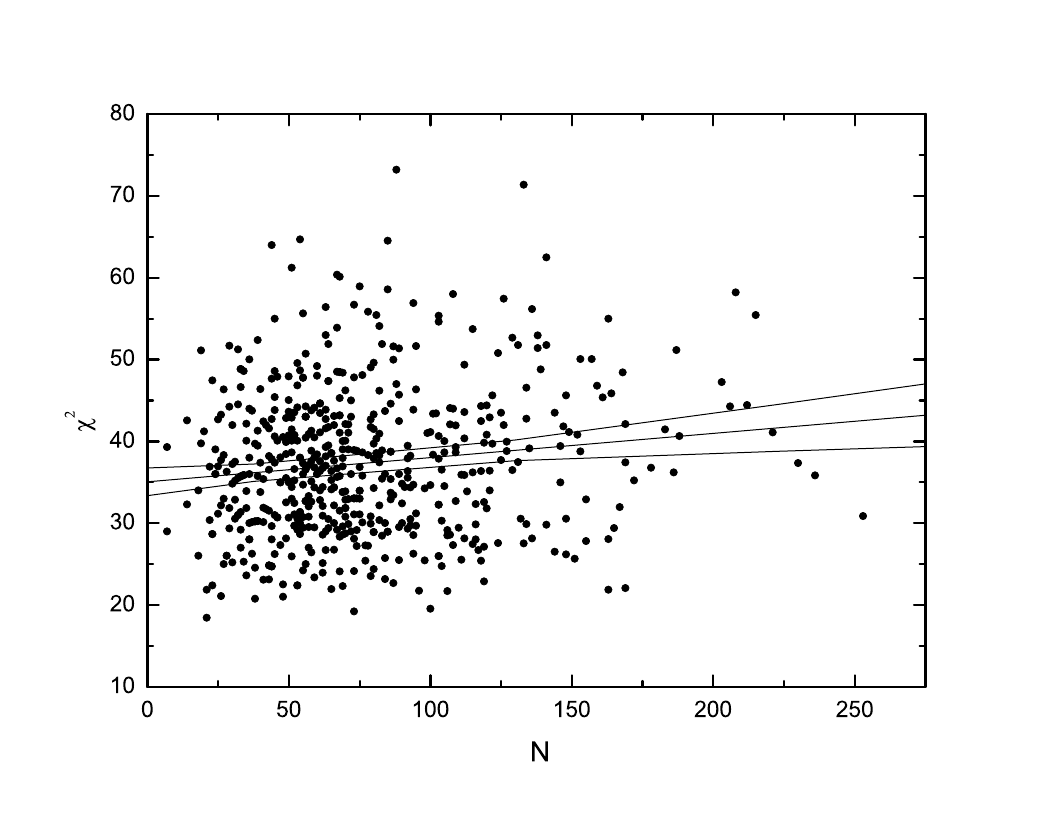}
		\label{fig:2}}
	\subfigure[Relation between $N$ and $\Delta_1/\sigma(\Delta_1)$]{\includegraphics*
		[width=0.45 \textwidth]{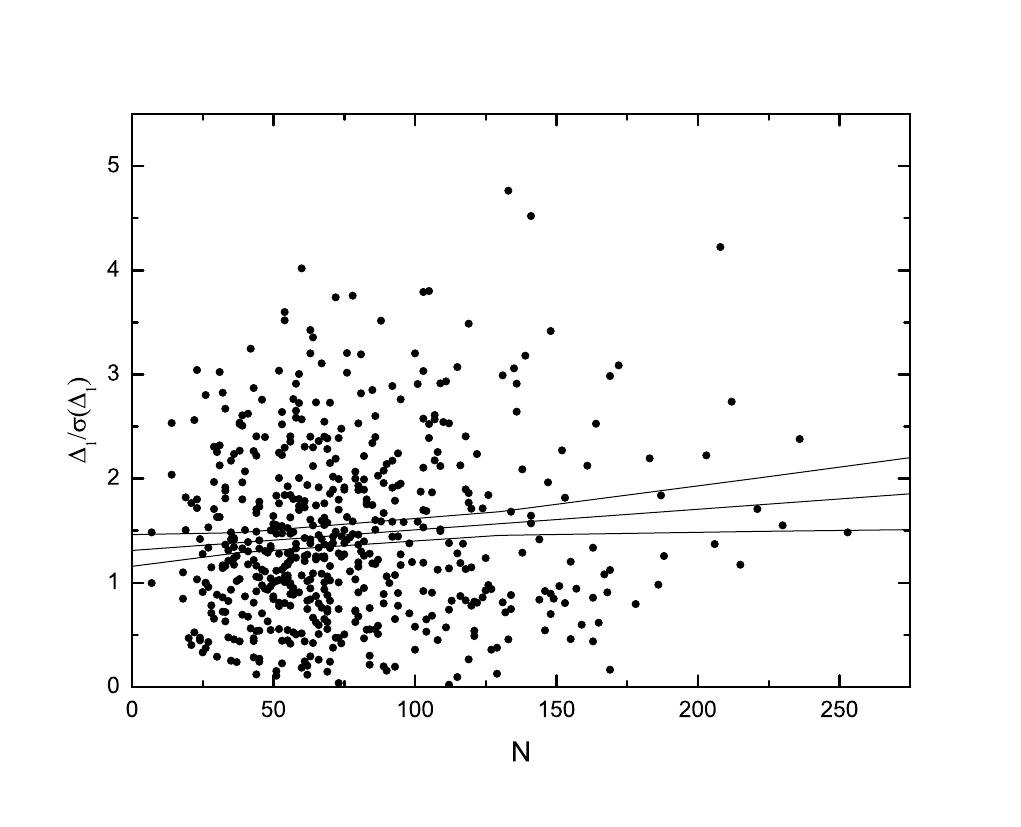}
		\label{fig:3}}
	\subfigure[Relation between $N$ and $\Delta/\sigma(\Delta)$]{\includegraphics*
		[width=0.45 \textwidth]{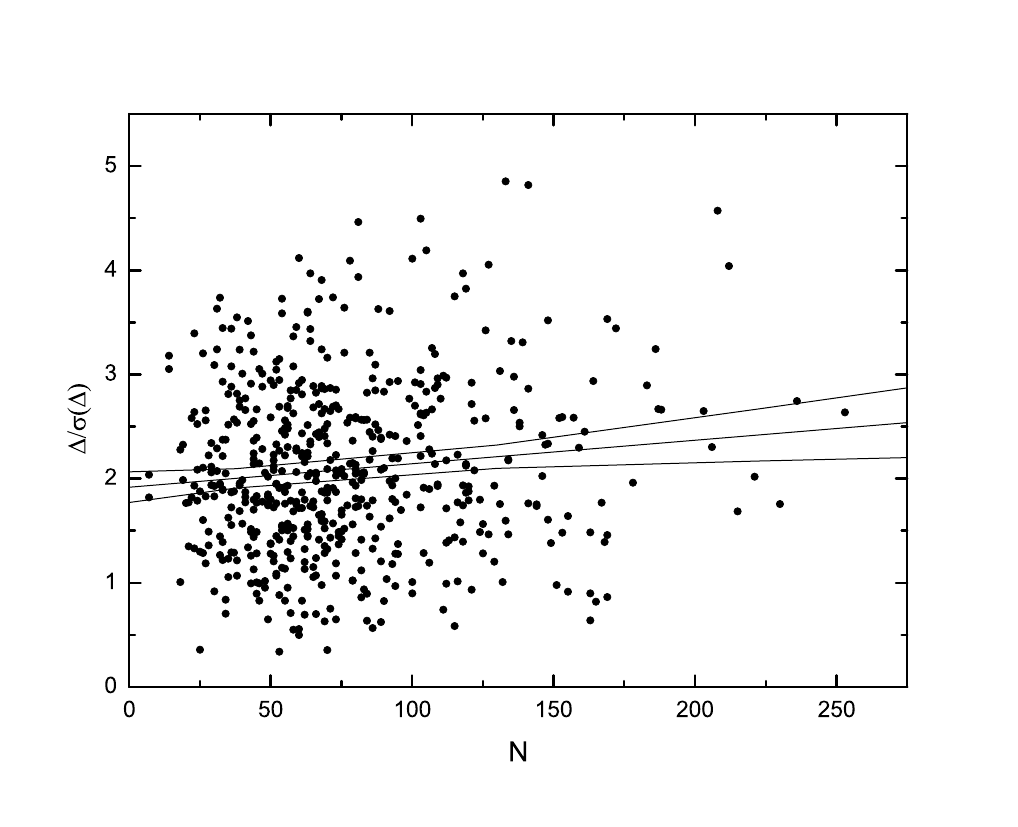}
		\label{fig:4}}
	\subfigure[Relation between $N$ and $C$]{\includegraphics*
		[width=0.45 \textwidth]{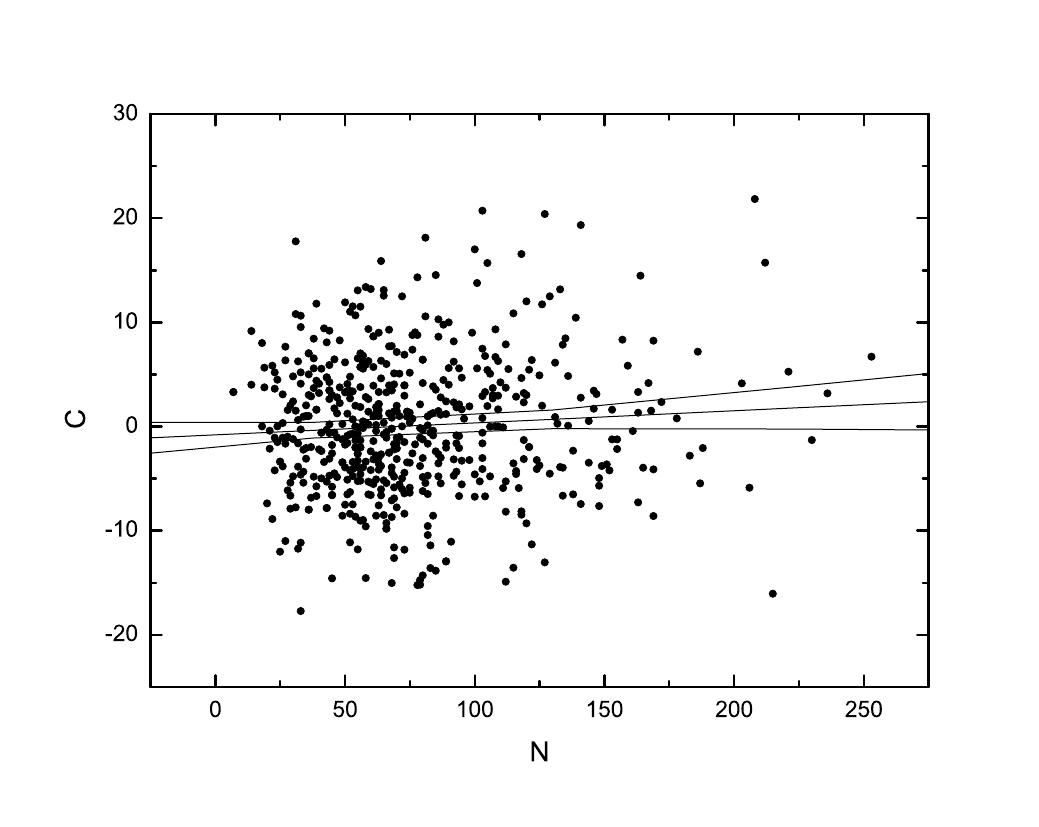}
		\label{fig:5}}
	\subfigure[Relation between $N$ and $\lambda$]{\includegraphics*
		[width=0.45 \textwidth]{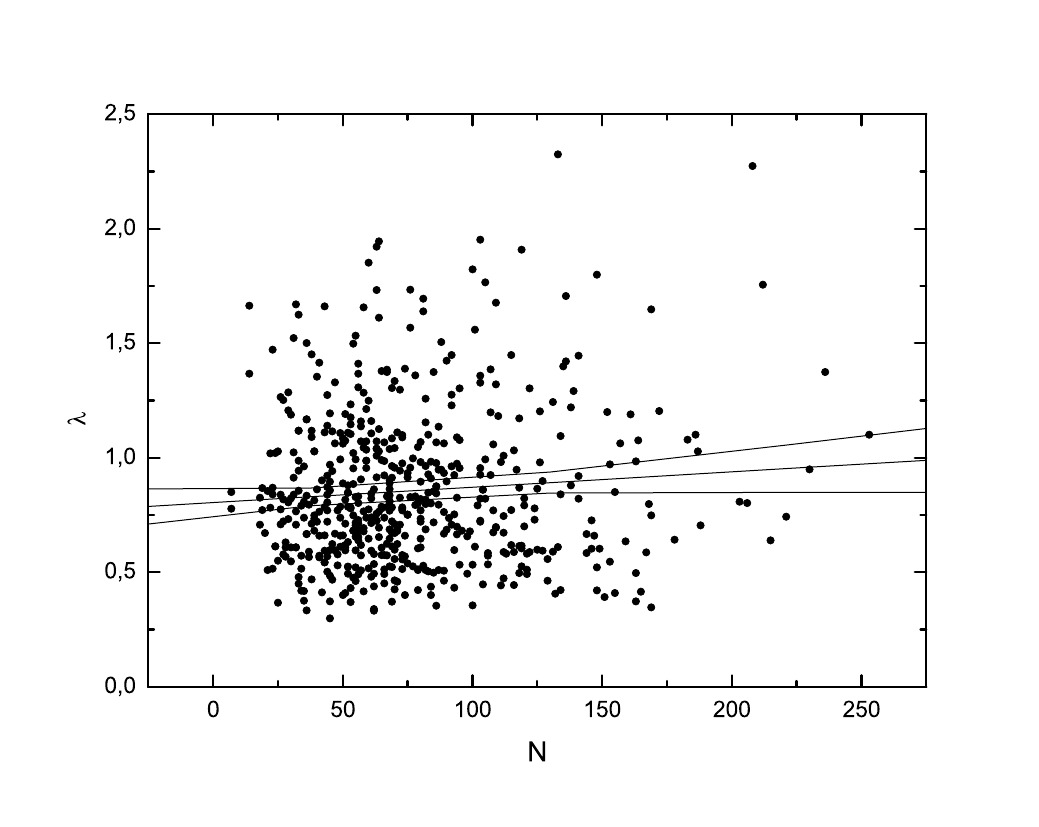}
		\label{fig:6}}
	\caption{The dependence of the number $N$ of galaxies in a cluster on different statistical parameters of the Sample B.
The bound errors, at the confidence level $95\%$, were presented as well.}\label{usz2}
\end{figure}

\begin{figure}
	\subfigure[Relation between redshift $z$ and $\chi^2$]{\includegraphics*
		[width=0.45 \textwidth]{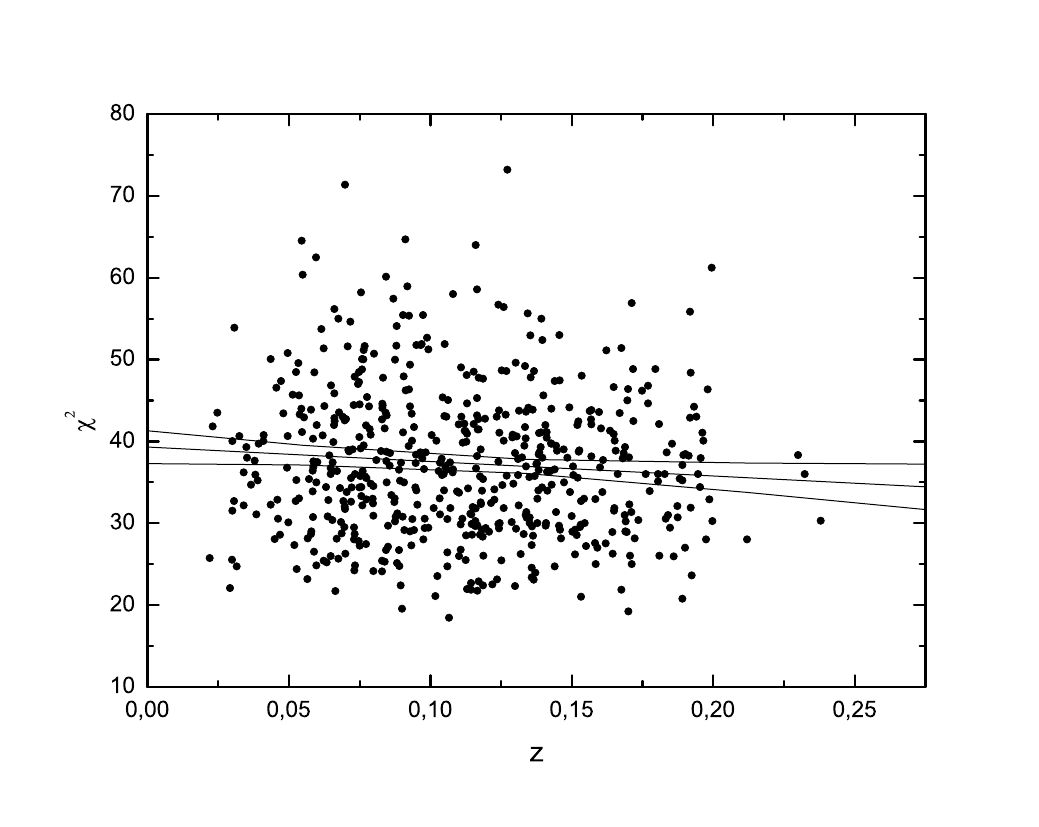}
		\label{fig:7}}
	\subfigure[Relation between redshift $z$ and $\Delta_1/\sigma(\Delta_1)$]{\includegraphics*
		[width=0.45 \textwidth]{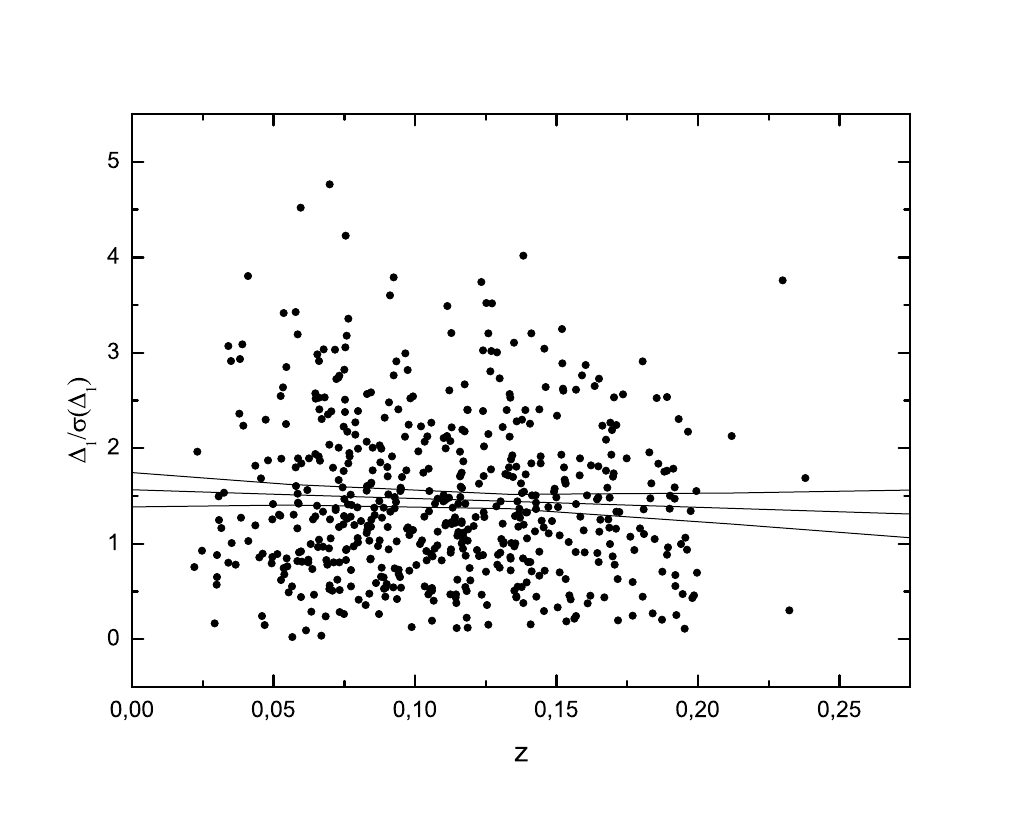}
		\label{fig:8}}
	\subfigure[Relation between redshift $z$ and $\Delta/\sigma(\Delta)$]{\includegraphics*
		[width=0.45 \textwidth]{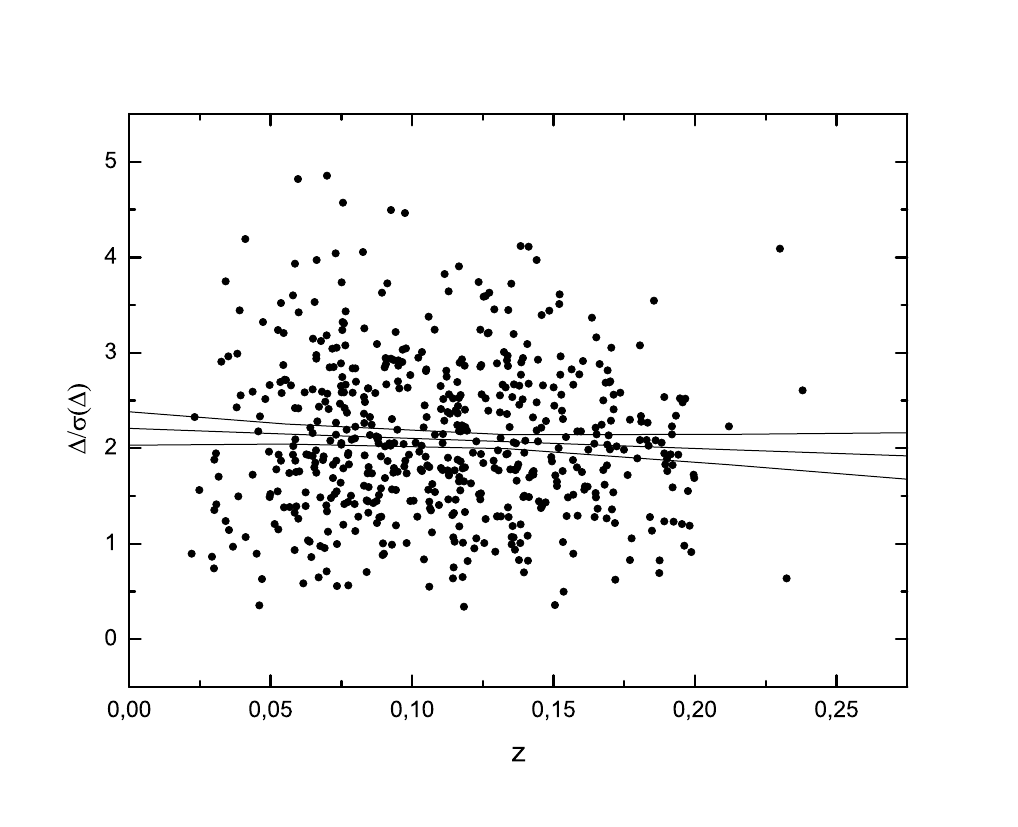}
		\label{fig:9}}
	\subfigure[Relation between redshift $z$ and $C$]{\includegraphics*
		[width=0.45 \textwidth]{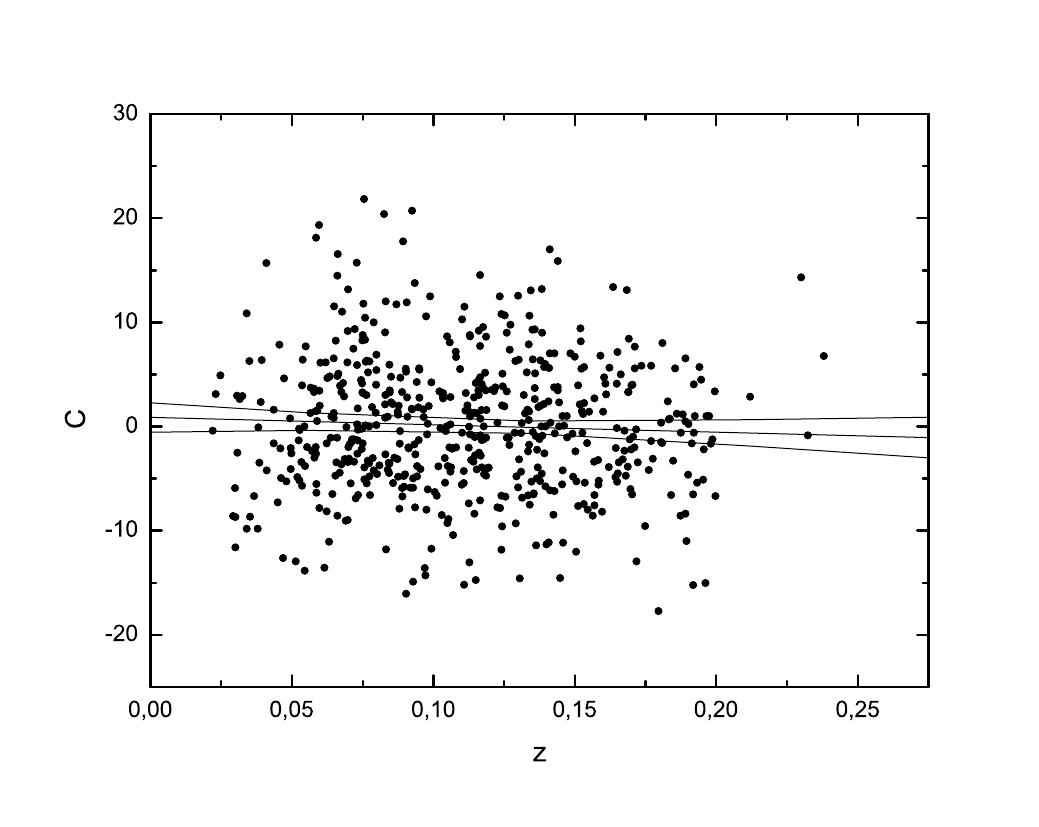}
		\label{fig:10}}
	\subfigure[Relation between redshift $z$ and $\lambda$]{\includegraphics*
		[width=0.45 \textwidth]{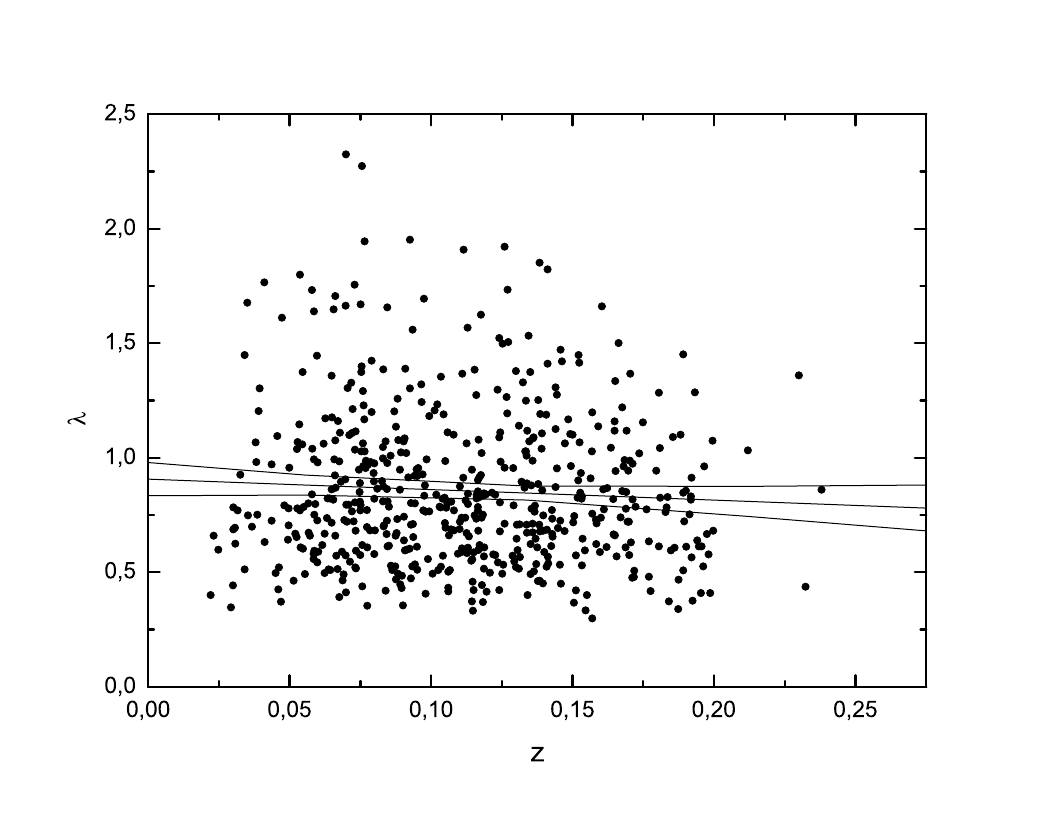}
		\label{fig:11}}
	\caption{The dependence of cluster redshift $z$ on different statistical parameters of the Sample B. The bound errors, 
at the confidence level $95\%$, were presented as well.}\label{usz3}
\end{figure}

\section{Observational Data}

First part of our data is the sample of rich Abell clusters containing at least 100 member galaxies each \citep{g19}.
The sample contains 247 clusters and it was selected on the base of the PF catalogue (\citet{Panko06}), see \citet{g19}
for details). However in the present paper we decide to restrict ourselves to 187 clusters which have explicit redshifts. 
As our PF cluster sample was not sufficient to confirm hypothesis that galaxies alignment decreases with redshift,
we decide to enlarge our sample on the DSS base.

From ACO Catalogue \citep{Abell89}  we selected all Abell clusters with galactic latitude $b>40^\circ$ and richness 
class $\geq1$. This gives us 1238 structures of galaxies from which we selected only those with redshifts $z<0.2$ 
\citep{Struble99}.  Therefore, 377 clusters have been left for analysis. From DSS we extracted the area covering 
2Mpc $\times$ 2Mpc ($h=0.75$, $q_0=0.5$) around each cluster. We applied the FOCAS package \citep{Jarvis81} to the 
extracted regions and obtained catalogues of galaxies, considering objects within the magnitude range ($m_3$, $m_3+3$), 
where $m_3$ is the magnitude of the third brightest galaxy. 
The catalogues obtained automatically were visually corrected in order to reduce the possible incorrect star/galaxy 
classification. FOCAS calculates the catalogue parameters using the moments of pixel distribution in an object. 
There are three steps from the basic image to the object list in FOCAS: segmentation, area assembly and object 
evaluation. The various parameters characterizing the individual images in the segmented areas were calculated.
In FOCAS, the object location is defined by centroids: 
\begin{equation}
\overline{x}=\frac{1}{M_{00}}\sum_{A}x_i\left[I(x,y)-I_s\right].
\end{equation}
\begin{equation}
\overline{y}=\frac{1}{M_{00}}\sum_{A}y_i\left[I(x,y)-I_s\right],
\end{equation}
where $M_{00}$ is zero moment, which is equal to:
\begin{equation}
M_{00}=\sum_{A}\left[I(x,y)-I_s\right].
\end{equation}
The summation over A means that the sum includes all pixels in the object-defining area A. $I(x, y)$ is the intensity 
corresponding to the density at the location $(x, y)$ in the digital plate image. $I_s$ is the intensity corresponding 
to the average plate density at the object location. Shape information about the object is obtained from the higher 
central moments:
\begin{equation}
M_{ij}=\sum A(x-\bar{x})^i(y-\bar{y})^j[I(x,y)-I_s].
\end{equation}

The object position angle is calculated using above central moments:
\begin{equation}
\tan(2\theta)=\frac{2M_{11}}{M_{20}-M_{02}}.
\end{equation}
Galaxy ellipticity reads
\begin{equation}
e=1-\frac{\lambda_2}{\lambda_1},
\end{equation}
where
\begin{eqnarray}
\lambda_1^2=\frac{1}{2}\left((M_{20}+M_{02})+\sqrt{(M_{20}-M_{02})^2+4M_{11}^2}\right), \nonumber \\
\lambda_2^2=\frac{1}{2}\left((M_{20}+M_{02})-\sqrt{(M_{20}-M_{02})^2+4M_{11}^2}\right).
\end{eqnarray}
Each catalogue contains information about the right ascension and declination of each galaxy, 
its coordinates $x$ and $y$ on the photographic plate, instrumental magnitude, object area, 
galaxy ellipticity and the position angle of the major axis of galaxy image. The equatorial 
galaxy coordinates for the epoch 2000 were computed according to the rectangular coordinates 
of DSS scans. We calculate the position angle and ellipticity of each galaxy cluster using 
the method described by  \citet{Carter80} which is also based on the first five moments 
of the observed distribution of galaxy coordinates $x_i$, $y_i$.

\section{Statistical studies}

\citet{h4} propose to analyze the distribution of galaxies angular momenta by that 
of the observed position angles of the  galactic image major axes. The direction of the 
angular momentum is then believed to be perpendicular to that of the major galaxy axis.
This means that in the original version of the method the face-on and nearly face-on galaxies must be 
excluded from the analysis. This method can also be extended for the studies of the spatial orientation of 
galaxy planes \citep{f4}. 

The idea of \citet{h4} is to use the statistical tests for investigation of the position angles distribution.
The higher value of statistics means greater deviation from isotropic distribution i.e. stronger
alignment of galaxies angular momenta in the analyzed structures. Since \citet{h4} paper this method has become 
the standard tool for searching of galactic alignments. Recent  improvement and revision of this method was 
presented in \citet{g19}.

In the present paper we follow the analysis from \citet{raa17}.
The entire range of  investigated angles was divided into $n$ bins. As the aim of the method is to  
detect non-random effect in the galaxies orientation, we first check if considered  distribution deviates 
from isotropic one. Following \citet{raa17}, in the present paper we use $\chi^2$ and Fourier tests. We also extend 
our analysis for first auto-correlation and Kolmogorov-Smirnov (K-S)  tests (\citet{h4,f4,g10a,g2011b} 
see \citet{g19} for last review). 

The statistics $\chi^2$ is:
\begin{equation} \label{eq:c1}
\chi^2 = \sum_{k = 1}^n \frac{(N_k -N\,p_k)^2}{N\,p_k}= \sum_{k = 1}^n \frac{(N_k -N_{0,k})^2}{N_{0,k}},
\end{equation}
where $p_k$ are probabilities that chosen galaxy falls into $k$-th bin, $N$ is the total number 
of galaxies in a sample (in a cluster in our case), $N_k$ is the number of galaxies within the $k$-th  
angular bin and $N_{0,k}=Np_k$ is the expected number of galaxies in the $k$-th bin. 
Note that the number of degrees of freedom of the $\chi^2$ test  is $n-1$,   mean value $E(\chi^2)=n-1$ while
 the variance $\sigma^2(\chi^2)=2(n-1)$. As in our analysis $n=36$, we obtain the values $E(\chi^2)=35$ while $\sigma^2(\chi^2)=70$, i.e. $\sigma(\chi^2)=8.367$.

The first auto-correlation test quantifies the correlations between galaxy numbers in neighboring angle bins. 
The statistics $C$ reads
 
\begin{equation}\label{eq:c2}
C= \sum_{k = 1}^n \frac{(N_k -N_{0,k})(N_{k+1} -N_{0,k+1})}
{\left[ N_{0,k} N_{0,k+1}\right]^{1/2}},
\end{equation}
where $N_{n+1}=N_1$. 
When, as in present paper, we analyze the distribution of the position angles, than all $N_{k,0}=N\,p_k$ are equal to each other and $E(C)\, =-1$ while $D(C)\approx n$ (i.e. $\sigma(C) \approx \sqrt{n}=6$), see \citet{g2011b,g19} for details.

If we assume that deviation from isotropy is a slowly varying function, we can use the Fourier test:
\begin{equation}
\label{eq:f9}
N_k = N_{0,k} (1+\Delta_{11} \cos{2 \theta_k} +\Delta_{21} \sin{2
\theta_k}+\Delta_{12} \cos{4 \theta_k}+\Delta_{22} \sin{4\theta_k}+.....).
\end{equation}
In this test, the statistically important are the amplitudes
\begin{equation}
\label{eq:f6}
\Delta_1 = \left( \Delta_{11}^2 + \Delta_{21}^2 \right)^{1/2},
\end{equation}
(only the first Fourier mode is taken into account) or 
\begin{equation}
\label{eq:f16}
\Delta =
 \left( \Delta_{11}^2 + \Delta_{21}^2+\Delta_{12}^2 + \Delta_{22}^2 \right)^{1/2},
\end{equation}
where the first and second Fourier modes are analyzed together. We investigate the   statistics $\Delta_1/\sigma(\Delta_1)=(\Delta^2_{11}/\sigma^2(\Delta_{11})+\Delta^2_{21}/\sigma^2(\Delta_{21}))^{1/2}$ 
and $\Delta/\sigma(\Delta)=(\Delta^2_{11}/\sigma^2(\Delta_{11})+\Delta^2_{21}/\sigma^2(\Delta_{21})+
\Delta^2_{12}/\sigma^2(\Delta_{12})+\Delta^2_{22}/\sigma^2(\Delta_{22}))^{1/2}$ 
(see \citet{g10a,g2011b,g19} for details). Note that the expression $\frac{\Delta^2_j}{\sigma^2(\Delta_j)}$ means
that elements of $\Delta^2$ should be divided by their errors rather then that the total factor $\Delta_j^2$ is divided by its error. Namely, the expectation values of the total factors are $E\left(\frac{\Delta_1}{\sigma(\Delta_1)}\right) =1.2247$ and $E\left(\frac{\Delta}{\sigma(\Delta)}\right)=1.8708$ while $\sigma^2(\Delta_{1}/\sigma(\Delta_{1}))=1/2$, and $\sigma^2(\Delta/\sigma(\Delta))=1/2$ (i.e. errors of the total factors equal to $\sqrt{2}$) - see \citet{g2011b,g19} for details.

In the case of K-S test, the statistics under study is $\lambda$:
\begin{equation}\label{eq:k1}
\lambda=\sqrt{N}\,D_n
\end{equation}
which is given by limiting Kolmogorov distribution, where
\begin{equation}
\label{eq:k2}
D_n= \sup|F(x)-S(x)|
\end{equation}
and $F(x)$ and $S(x)$ are theoretical and observational distributions of the 
investigated angle respectively. \citet{Wang03} analyzing the limiting form of $D_n$ function found that $\mu(\lambda)=0.868731$ while $\sigma^2(\lambda)=0.067773$ i.e $\sigma(\lambda)=0.260333$ (see also \citet{g19} for discussion).

Using extended \citet{h4} method it is possible to analyze both the alignment dependence on particular parameter 
like richness of galaxy cluster \citet{g10a} and quantitatively answer the question if an alignment is present 
in a sample (see \citet{g19} for last revision). In our previous papers \citet{g10a,raa17}, using sample of 
247 PF rich Abell clusters,  it was shown that alignment of galaxies in a cluster increases significantly 
with its richness. Unfortunately available data was insufficient for persuasive conclusion about correctness of theoretically 
predicted by \citet{apj15,raa17} dependence of analyzed statistics on redshift $z$. 

For this reason we  perform the investigations of our samples of galaxy clusters checking if there is a significant
dependence between analyzed statistics and both richness and redshifts of the clusters. As a first step we analyzed 
the linear model  $Y=aX+b$. The $Y$ are the values of analyzed statistics i.e. 
$\chi^2$, $\Delta_1/\sigma(\Delta_1)$, $\Delta/\sigma(\Delta)$, $C$ and $\lambda$ 
(see \citet{raa17} for details) while $X$ is the number of analyzed galaxies in each particular cluster
or its redshift $z$ respectively. Our null hypothesis $H_0$  is that analyzed statistics $Y$ does not depend
on $X$ .  This means that we should analyze the statistics  $t=a/\sigma(a)$, which has Student's  distribution 
with $u-2$ degrees of freedom, where $u$ is the number of analyzed clusters. In other words, we test $H_0$
hypothesis that $t<0$ against $H_1$ hypothesis that $t>0$, where $t>0$ corresponds to the case of dependence on 
number of member galaxies in clusters and $t<0$ to the case of dependence on redshift $z$. In order to reject
the $H_0$ hypothesis, the value of observed statistics $t$ should be
greater than $t_{cr}$ which could be obtained from the tables. For example, for sample of 247 clusters 
analyzed in \citet{raa17} (our sample A) at the significance level $\alpha=0.05$,  the value $t_{cr} = 1.651$. For our sample 
B (564 clusters) at the significance level $\alpha=0.05$,  the value $t_{cr} = 1.648$.

The results of our  analysis are presented in the table \ref{tab:t1} and figs \ref{usz2},\ref{usz3}.
One could discern from the table \ref{tab:t1}, that if the analysis of the sample A confirms the alignment increasing with cluster richness, then any test confirm negative deviation of linear regression parameter from zero in the case of alignment dependence 
on redshift. However, both above dependencies can be confirmed from the full sample (564 clusters) analysis. At first we 
analyzed statistics i.e. alignment, which increases significantly with richness of a cluster, 
confirming the result obtained in \citet{g10a, raa17} as well as theoretical prediction \citet{apj15}. 
The details are also presented in the figures \ref{fig:2} - \ref{fig:6}. Moreover, we could conclude that 
alignment decreases with $z$, which means that it increases with time as predicted by \citet{apj15,raa17,Schmitz18} 
(see figures \ref{fig:7} - \ref{fig:11}).

However, closer look at the results show that situation is not so clear yet. This is because in real data the cluster richness usually decreases also with the redshift $z$. Quantitatively, in linear model, the dependence between 
richness and redshift of a cluster is $N(z)=az+b$. In this model, we obtain the value of $t$ statistics $t=-7.066$. 
This is the reason that we repeated our analysis as 3D model $Y=a_1N+a_2z+b$. Note, that until now, such 3D analysis 
has not been performed in galaxies alignment studies, but due to the above reason, we consider it to be necessary here. 
In this extended analysis, the test statistics $t$ are given by formulae  $t_1=a_1/\sigma(a_1)$ and $t_2=a_2/\sigma(a_2)$. 
From Table 2, we could confirm that alignment increases significantly with cluster richness. One should note that all 
our tests show that statistics are decreasing with $z$, however values of $t$ statistics are too small to make 
statistically significant (significance level $\alpha=0.05$) effect.  

\begin{table}[t]
	\begin{center}
		\caption{The  statistics $t=a/\sigma(a)$ for our sample of Abell clusters.
			Sample A - 247 rich Abell clusters from PF catalogue (as in Stephanovich \& God{\l}owski (2017)).
			Sample B - full sample of 564 clusters (directly known redshift)
		}
		\label{tab:t1}
		\begin{tabular}{ccc}
			Test&$S=f(N)$&$S=f(z)$\\
			Sample $A$                       &       &       \\                                                   
			$\chi^2$                         &$1.872$&$-0.769$\\                        
			$\Delta_{1}/\sigma(\Delta_{1})$  &$1.613$&$ 0.611$\\                      
			$\Delta/\sigma(\Delta)$          &$1.964$&$-0.066$\\                      
			$C$                              &$1.352$&$ 1.343$\\                     
			$\lambda$                        &$2.366$&$ 0.176$\\                              
			Sample $B$                       &       &       \\                              
			$\chi^2$                         &$3.402$&$-2.342$\\                      
			$\Delta_{1}/\sigma(\Delta_{1})$  &$2.857$&$-1.452$\\                      
			$\Delta/\sigma(\Delta)$          &$3.142$&$-1.646$\\			     
			$C$                              &$1.825$&$-1.305$\\                      
			$\lambda$                        &$2.333$&$-1.953$\\                      
		\end{tabular}
	\end{center}
\end{table}

\begin{table}[t]
\begin{center}
\caption{The  statistics $t=a/\sigma(a)$ for 3D analysis of our sample of Abell clusters.
Sample B - full sample of 564 clusters (directly known redshift)}
\label{tab:t2}
\begin{tabular}{ccc}
Test&$S=f(N)$&$S=f(z)$\\
                                                                       
Sample $B$                       &       &       \\                              
$\chi^2$                         &$2.846$&$-1.434$\\                      
$\Delta_{1}/\sigma(\Delta_{1})$  &$2.250$&$-0.718$\\                      
$\Delta/\sigma(\Delta)$          &$2.625$&$-0.811$\\			     
$C$                              &$1.538$&$-0.814$\\                      
$\lambda$                        &$2.000$&$-1.302$\\                      %
\end{tabular}
\end{center}
\end{table}

\begin{table}[t]
\begin{center}
\caption{The Ilustris relation between Angular Momentum and mass simulations}
\begin{tabular}{cccc}
Mass & $a$ & $\sigma(a)$&$t=a/\sigma(a)$\\
$>10^{12}$   &   1.807  &    0.028   &   66.93\\
$>10^{13}$   &   1.708   &   0.114  &    14.94 \\
\end{tabular}
\end{center}
\end{table}

\section{Simulations}

The \citet{Illustris} was the simulation base for our present study. The Project uses the AREPO code for hydrodynamic 
realizations of a $(106.5 Mpc)^3$ cosmological volume \citep{Springel10}. The simulation assumes a $\Lambda CDM$ 
cosmology with the 
$\Omega_m = 0.2726$, $\Omega_\Lambda = 0.7274$, $\Omega_b = 0.0456$, $\sigma_8 = 0.809$, $n_s = 0.963$, 
and $H_0 = 100\cdot h\cdot km \cdot s^{-1} Mpc^{-1}$ with $h = 0.704$.   It contains multiple 
resolution runs with the highest resolution performed for Illustris - 1. Three different physical 
configurations have been applied: dark matter only as well as non-radiative and full galaxy formation.
In the first case (dark matter only), the mass was treated as collisionless in the simulations.
The non-radiation configuration also adds gas hydrodynamics, but ignores radiative cooling and star formation processes. 
The full galaxy formation physics contains (in addition to the previously mentioned ones) also processes related to 
galaxy emergence through a model described in \citet{Vogelsberger13}. Illustris-1 consists of $136$ runs for different 
redshifts $z$ where the initial conditions was generated at $z=127$ for snapshot $0$ and evolved to $z=0$ for snapshot $135$. 

Illustris successfully follows the coevolution of dark and visible matter. Haloes, subhaloes, and their basic properties 
have been identified with the FOF and SUBFIND algorithms \citep{Davis85,Springel01,Dolag09}, at every of the 136 stored 
snapshots. We have added information from the supplementary catalog to the resulting directory of Haloes from \citet{Zjupa17}.
The code was written in such a manner that it can run both as a postprocessing option to increase existing catalogues or 
as part of the regular group finding.

From Illustris-1 we select Haloes at $z = 0$. We obtained 119 Haloes with total mass exceeding $10^{13}M_{\odot}$  
and 1435 ones with total mass higher than   $10^{12}M_{\odot}$.   The angular momentum parameter for extracted 
Haloes was taken from \citet{Zjupa17}.
 
Ilustris simulations give direct value of both mass of the structures and their angular momentum. 
Present available data from Ilustris are evolved to $z=0$, so it is possible to study the dependence of angular
momentum as the function of cluster mass but unfortunately not of redshift.
However as we know directly the cluster angular momentum, it is not necessary to assume the linear relation between 
angular momentum and mass. Since theoretical modeling predicts usually the power law relations (see also \citet{apj15} 
for review) we could study the model $J=b\cdot M^a$. The latter relation could easily be rendered as a linear model 
$\ln\,J=\ln\, b+a\ \ln\, M$. 

\begin{figure}[t]
	\subfigure[$M>10^{12}M_{\odot}$]{\includegraphics*
		[width=0.5 \textwidth]{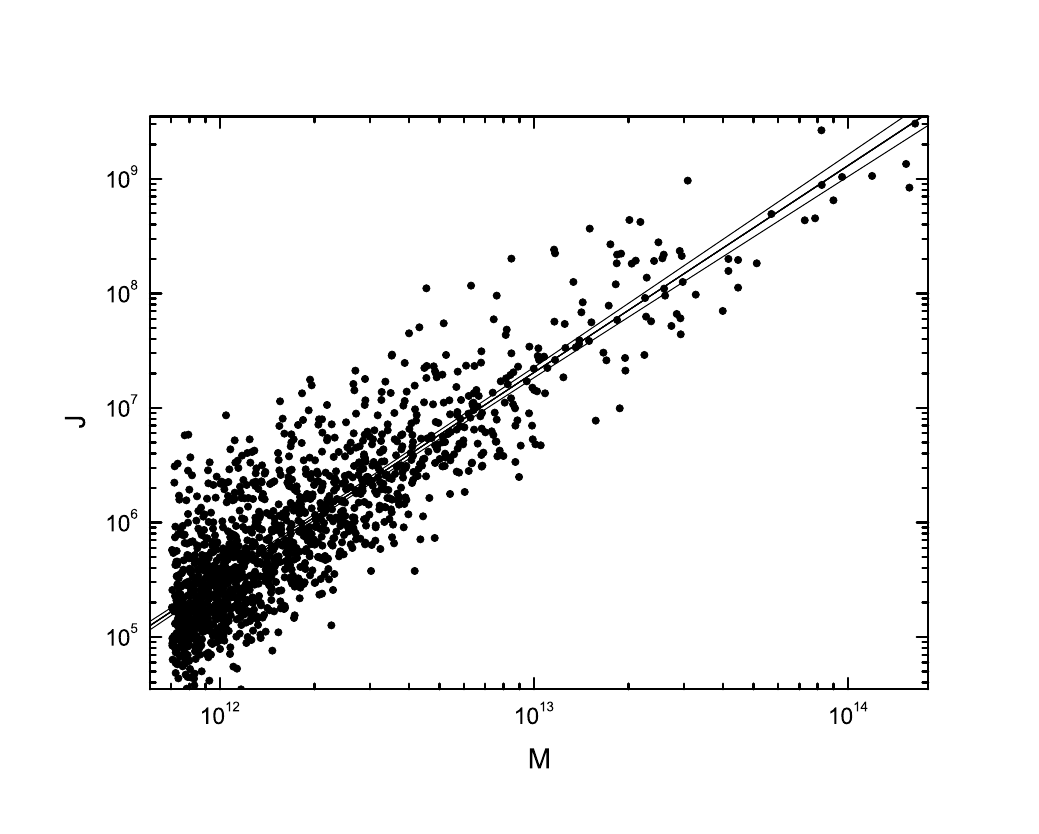}
		\label{fig:12}}
	\hfill
	\subfigure[$M>10^{13}M_{\odot}$]{\includegraphics*
		[width=0.5 \textwidth]{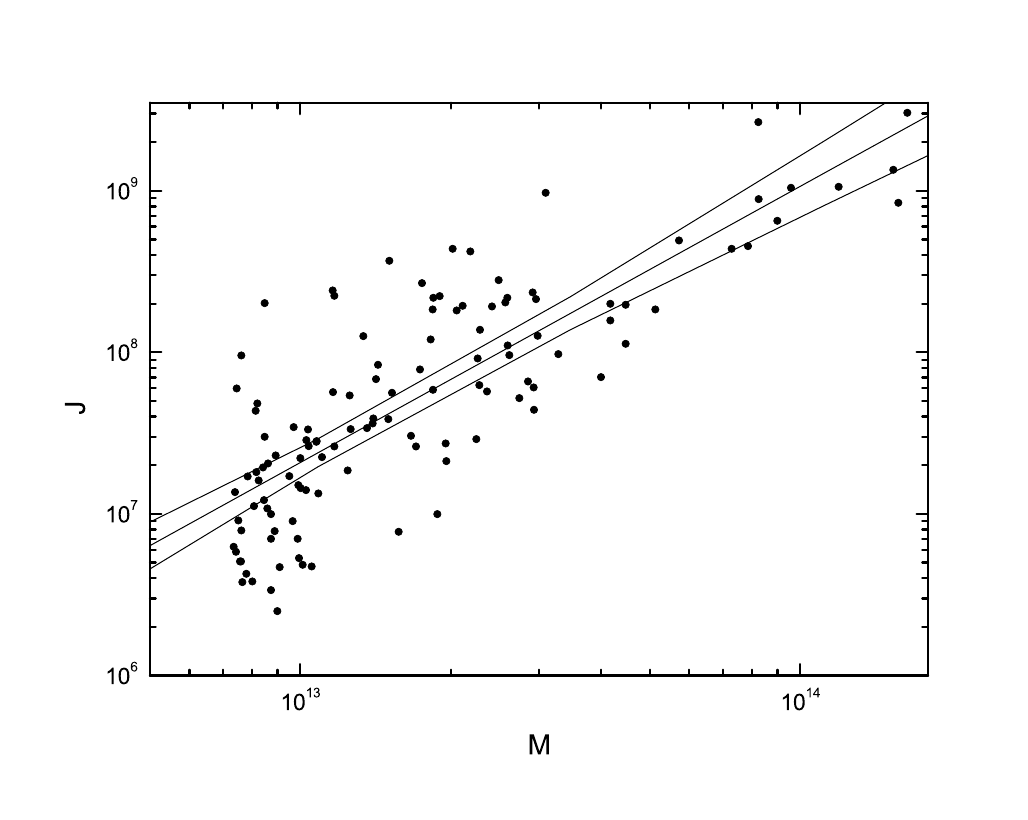}
		\label{fig:13}}
	\caption{Relation between angular momentum and mass of a cluster derived from Ilustris simulation for different total 
masses $M$. The bound errors, at the confidence level $95\%$, were presented as well.}\label{usz4}
\end{figure}

The results of the analysis are presented in the table 3 and figure \ref{usz4}. Analysis of Ilustris Simulation confirms 
that angular momentum of a cluster increases with its mass. In this case, the coefficient $a=1.807 \pm 0.028$ (fig. \ref{fig:12}) 
The analysis of a sample, where only the most massive clusters (mass $M>10^{13}$ solar mass) are left, gives $a=1.708 \pm 0.114$ (fig. \ref{fig:13}) which is close to most popular theoretical prediction $a=5/3 \approx 1.667$ (see \citet{g10a,apj15,raa17} for details). 

\section{Outlook}

In the present paper we have shown that the distribution of masses of the stellar objects does not alter substantially the 
distribution function of their gravitational fields. This shows that the main contribution to latter distribution function 
comes from the long-range Newtonian interaction between astronomical objects rather than from the distribution of their masses. At the same time, the mass distribution alters the dependence of $L_{max}$ on astronomical objects concentration $n$ (see Eq. \eqref{irs1})  and total cluster mass $M$ (see Eq. \eqref{irs2}), which is observationally important.  
To discern, which of the dependencies \eqref{irs1} or \eqref{irs2} is realized in practice, the additional observational 
data is needed. As total interaction potential contains both luminous and dark matter components, one can ask a question about alignment of sub-dominant galaxies, even though the majority of galaxy clusters angular momenta is related to the smooth dark matter halo component. This question becomes important in view of the fact that mass distribution \eqref{yt1} alters the dependence on total cluster mass $M$ \eqref{irs2}. Namely, in the halo model \citep{sb10}, where the galaxies are embedded in a dark matter halo, latter may mediate the intergalactic interaction, adding possible short-range terms to it. That is to say, to "see each other" in a dark matter halo, the galaxies should go closer than in an empty space. 

Note, that the observational results about lack of alignment of galaxies for less clumpy (so called poor) clusters, as well 
as evidence for such alignment in the clumpy (rich) ones (\citet{Godlowski05,Aryal07},  see also \citet{g2011a} for incremental 
study and relevant references) clearly shows that angular momentum of galaxy groups and clusters increases with their mass 
(richness).  The generalized analysis, based on Eq. \eqref{sa2}, where $n=n(r,m)$ (i.e. the mass becomes  spatially distributed), will improve the overall understanding, which can additionally be tested against observed galaxy shape distributions and alignments. The problem of angular momenta alignment due to their interactions as well as those with dark matter haloes has been simulated by \citet{Hahn07}. The main effect there is the presence of a threshold cluster mass (richness) value. Latter is related to mutual alignment of clusters and dark matter haloes axes. This fact can be analyzed on the base of more general model \eqref{sa2}, which accounts for spatially inhomogeneous distribution of number density of stellar objects as well as for its mass dependence. We postpone the consideration of this interesting question for the future publications.

Our formalism permits  studying this effect (see \citet{raa17}) as well as the nonequilibrium time evolution of 
luminous astronomical objects  (with respect to dark matter haloes) within the $\Lambda$CDM model. The combination of 
stochastic dynamical approaches  \citep{gs1,gs2} along with deterministic one defined in $\Lambda$CDM model, 
may permit to answer (at least qualitatively) the question about the galaxies (and their clusters) initial 
alignment at the time of their formation. The questions about how dark matter haloes influence (mediate) 
latter alignment, can also be answered within the above dynamic approach. 

Our statistical analysis of Abell clusters sample shows that alignment of galaxies and their clusters angular momenta increases substantially with cluster mass. This result is confirmed also by 3D analysis, consisting in the studies of the
 dependence of galaxies alignment in a cluster both on its richness and redshift. We have also found 
that alignment decreases with redshift i.e. increases with time, but above 3D studies show that this effect is too faint to be 
confirmed statistically at the significance level $\alpha=0.05$. Probable reason is that the corresponding relaxation time is 
too long. So for future investigations more extended data containing larger number of galaxy clusters (and with higher redshifts) 
are required. The comparison of our theoretical results with those of the Ilustris simulation also confirms the increase of 
galaxies angular momenta with cluster mass. Moreover, latter comparison confirms the power law relation with coefficient, very close to 5/3, which is the value favored by most popular theoretical predictions - see, for instance, \citet{apj15,raa17}. 
This may point that the approach of \citet{apj15,raa17} correctly reflects main features of galaxies and their clusters formation. 

Let us finally note, that as we have shown above, our results are in close conformity with commonly preferred model of the 
galaxies formation, i.e. so-called hierarchic clustering model \citep{peb69}, improved recently by taking into account tidal 
torque scenario.

\end{document}